\documentclass[preprint,12pt]{elsarticle}

\usepackage{amssymb}
\begin{document}

\begin{frontmatter}

%% Title, authors and addresses

%% use the tnoteref command within \title for footnotes;
%% use the tnotetext command for theassociated footnote;
%% use the fnref command within \author or \address for footnotes;
%% use the fntext command for theassociated footnote;
%% use the corref command within \author for corresponding author footnotes;
%% use the cortext command for theassociated footnote;
%% use the ead command for the email address,
%% and the form \ead[url] for the home page:
%% \title{Title\tnoteref{label1}}
%% \tnotetext[label1]{}
%% \author{Name\corref{cor1}\fnref{label2}}
%% \ead{email address}
%% \ead[url]{home page}
%% \fntext[label2]{}
%% \cortext[cor1]{}
%% \address{Address\fnref{label3}}
%% \fntext[label3]{}

\title{Epidemic spreading driven by biased random walks}

%% use optional labels to link authors explicitly to addresses:
%% \author[label1,label2]{}
%% \address[label1]{}
%% \address[label2]{}

\author{Cunlai Pu, Siyuan Li, Jian Yang}

\address{Nanjing University of Science and Technology, Nanjing 210094, China}

\begin{abstract}
%% Text of abstract
Random walk is one of the basic mechanisms found in many network applications. We study the epidemic spreading dynamics driven by biased random walks on complex networks. In our epidemic model, each time  infected nodes  constantly spread some infected packets by biased  random walks to their neighbor nodes causing the infection of the susceptible nodes that receive the packets.  An infected node get recovered from infection  with a fixed probability. Simulation and analytical results on model and real-world networks show that the epidemic spreading becomes intense and wide  with the increase of delivery capacity of infected nodes, average node degree, homogeneity of node degree distribution.  Furthermore, there are corresponding optimal parameters such that  the infected nodes have  instantaneously the largest population, and the epidemic spreading process covers the largest part of a network.
\end{abstract}

\begin{keyword}
%% keywords here, in the form: keyword \sep keyword
epidemic spreading \sep biased random walks \sep complex networks
%% PACS codes here, in the form: \PACS code \sep code

%% MSC codes here, in the form: \MSC code \sep code
%% or \MSC[2008] code \sep code (2000 is the default)

\end{keyword}

\end{frontmatter}

%% \linenumbers

%% main text
\section{Introduction}
Unexpected outbreak of many  epidemics in biological systems\cite{Green,Hethcote} and the spread of viruses in technology systems\cite{Balthrop,WangP,vesp} result in a lot of death or great damage in related systems. The study of epidemiological models
has a long history, especially in the field of social science\cite{Brauer,Berkman}. The SIR (susceptible-infected-removed) model and the SIS (susceptible-infected-susceptible) model are two representative models which capture the basic properties of epidemic spreading through the transition among several  disease states\cite{bailey,May}. In SIR model, a susceptible individual will become infected with certain rate when it has contact with infected individuals. An infected individual will get immunity to the disease or die at some constant rate, and becomes a removed node  which means it can not get infected again. Therefore, the spread of disease will terminate when all the infected individuals are removed from the disease. Differently in SIS model, there are just two states, susceptible and infected.  A recovered individual can get infected again.   If the fraction of infected individuals is large enough the disease will spread indefinitely, otherwise it will die out after sometime. Initially, the epidemic models are considered under the homogeneous mixing hypothesis\cite{Anderson}, in which it assumes that each time  an arbitrary individual has an equal opportunity to contact with everyone else in the population.  Later, results from network science community demonstrate that most real-world networked systems have heterogeneous topological structures\cite{Newman09,Dor08,Barrat}, and this greatly promote many mathematicians and physicists to explore  epidemic models on heterogeneous random networks\cite{Taoz,Rui,Gang,Shenwe} by means of mean-field approximation\cite{pastor,Yangz,sahneh}, generating functions formalism\cite{MEJNEWMAN} and percolation theory\cite{Cohenr}. It was found that for random networks with strongly heterogeneous degree distribution, like many real-world networks, the epidemic threshold is absent, which means epidemics always have a finite probability to survive indefinitely\cite{Newman09,pastor}.

Besides of topological properities,  traffics in networks also have great impacts on the epidemic spreading. For instance, in the Internet computer viruses transmit from a node to another one with data packets. Without transmission of  packets, viruses can not spread even if the two nodes are physically connected. Another example is the air traffics speed the spread of disease among different spatial areas.
The combination between epidemic spread and traffic dynamics were first considered in the metapopulation model\cite{colizza} which characterizes the dynamics of systems composed of subpopulations. Then, Meloni et al\cite{Meloni} studied the impact of traffic dynamics on the spread of virus in the Internet, in which the information packets are transmitted with the shortest path protocol. Later, many mechanisms were proposed to suppress the traffic-driven epidemic spreading, for instance controlling the traffic flow\cite{bajardi}, the routing strategy\cite{hxyang11,Yanghx14}, or the heterogeneous curing rate\cite{cshen}, deleting some particular edges\cite{yanghx13}, etc.

Random walk is one of the basic mechanisms related to spreading processes\cite{lovasz,Noh,Bonaventura}. For example, a mobile phone virus may randomly dial some phone numbers from the directory.  Some computer viruses propagate  randomly   by email or other online communication tools. Therefore, the role of random walks in the epidemic spreading should be explored.  We propose an epidemic model driven by biased random walks. In our model, an infected node sends infected packets at a constant rate  to its neighbor nodes through biased random walks. A susceptible node gets infected after receive the infected packets, and will be removed from the set of infected nodes with a constant rate. We inverstigate the spreading dynamics, the optimal control parameter of our model and the  influence of network topologies on our model. 
\section{SIR model}
We improve the traditional SIR model by incorporating the traffic dynamics driven by random walks.  
The SIR model is one of the traditional epidemic spread models in literature. In the SIR model, there are three types of nodes including susceptible nodes, infected nodes and removed nodes. A susceptible node is susceptible to epidemics. An infected node is already infected by the epidemic. A recovered node is the one that is removed from the set of infected nodes.  Assume the numbers of susceptible, infected and removed nodes at time $t $ are denoted as $S(t)$,  $I(t)$ and $R(t)$ respectively. There are three basic elements in the SIR model as follows\cite{bailey}: 
\begin{enumerate}
\renewcommand{\labelenumi}{(\theenumi)}
\item	Assume the number of nodes in the network is fixed to $N$. Then $N=S(t)+I(t)+R(t)$ for all $t$.
\item At time $t$,  an arbitrary infected node infects the susceptible nodes by a ratio $\beta$.  Then the increased number of infected nodes at $ t$  is $\beta \ast s(t) \ast I(t)$.
\item At time t, the number of infected nodes removed is proportional to the total number of infected nodes $ I(t)$, which is $ \lambda \ast I(t)$.
\end{enumerate}
According to the three elements, the dynamics of the SIR model can be expressed as follows\cite{bailey}:
\begin{equation}\label{pb1}
\left \{
\begin{array}{rl}
\frac{\mathrm{d}I}{\mathrm{d}t} &= \beta SI-\lambda I,\\
\frac{\mathrm{d}S}{\mathrm{d}t} &= -\beta S I,\\
\frac{\mathrm{d}R}{\mathrm{d}t} &= \lambda I.\\
   \end{array}
\right.
\end{equation}
When time $t$ is large enough,  all the infected nodes will eventually  becomes removed nodes, and the epidemic spreading stops.

The SIR model is based on the assumption that a node has equal probability to contact with every other node in a network. However, in real situations, individuals often have heterogeneous numbers of contacts\cite{Newman09}. A few individuals have large number of contacts which will get more contacts according to the “rich-get-richer” mechanism, while most of the individuals have a few contacts. Especially in the Internet, the epidemic can not spread from a node to another node unless there is transport of infected packets between the two nodes. Additionally, in city networks even the cities are physically well connected, an epidemic can not spread among cities unless individuals who get infected by the disease move among the cities. 
Therefore, epidemics are often correlated with traffics for their spreading. 

\section{Epidemic model driven by biased random walks}
We consider the traffic dynamics driven by biased random walks in the epidemic spreading process.  In our model, each time an infected nodes will delivery constantly $C $ infected packets to its neighbor nodes through biased random walks.  If an infected or removed node receives the packets, it will drop the packets. If a susceptible node receives the packet, it becomes an infected node and starts delivering infected packets from next time step. An infected node has the probability $ \lambda$ to become a removed node. The transitions among susceptible, infected and removed nodes are shown in Fig. 1.
\begin{figure}
 \centering
\includegraphics[width=5in,height=1in]{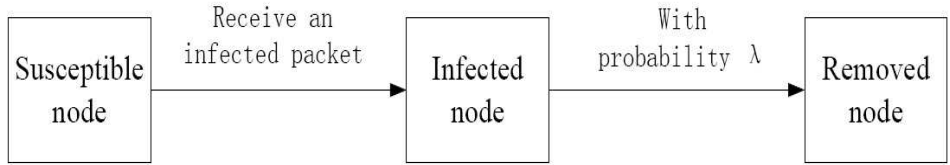}
\caption{Illustration of the transitions among the susceptible, infected and removed nodes in our model.}
\end{figure}
\subsection{Dynamics of our model}
Assume an arbitrary infected node $a$ which has $\langle K \rangle$ neighbor nodes.  $\langle K \rangle $ is the average node degree of the network. The degrees of $a$'s neighbor nodes are $ k_1,  k_2, \cdots, k_{\langle K \rangle}$  respectively. According to the biased random walk mechanism, for a neighbor node $i$ the probability that node $a$ sends an infected packet to node $i$ is as follows:
\begin{eqnarray}
P_{ai}=\frac{k_{i}^{\alpha}}{\sum_{j=1}^{\langle K \rangle}k_{j}^{\alpha}}.
\end{eqnarray}
Where $\alpha$ is the control parameter of the biased random walk. When $\alpha=0$, all the neighbor nodes have equal opportunity to receive an infected packet delivered from node $a$ which means they have equal probability to get infected. When $\alpha > 0$, nodes of larger degree have larger probability to receive the packet. When $\alpha < 0$, nodes of smaller degree have larger probability to receive the packet.
Since $a$ send $C$ infected packets each time, the probability that node $i$ will not receive any packets from node $a$ is:
\begin{eqnarray}
\bar{P}_{ai}=(1-\frac{k_{i}^{\alpha}}{\sum_{j=1}^{\langle K \rangle}k_{j}^{\alpha}})^{C}.
\end{eqnarray}
Assume $ X_i$ is an random variable that represents the event that node $i$ is infected. Then $ X_{i}=0$ means  node $i$ hasn't  get any infected packet from node $a$, and node $i$ is not infected.   $ X_{i}=1$ means  node $i$  has received at least one of the infected packets from node $a$, and node $i$ is  infected.  Then  we have:
\begin{equation}  
\left\{  
             \begin{array}{lr}  
             P(X_i=0)=(1-\frac{k_{i}^{\alpha}}{\sum_{j=1}^{\langle K \rangle}k_{j}^{\alpha}})^{C}, &  \\  
             P(X_i=1)=1-(1-\frac{k_{i}^{\alpha}}{\sum_{j=1}^{\langle K \rangle}k_{j}^{\alpha}})^{C}.&  
             \end{array}  
\right.  
\end{equation}  
Then the expected value of $X_i$ is:
\begin{eqnarray}
\mathrm{E}(X_i)=1-(1-\frac{k_{i}^{\alpha}}{\sum_{j=1}^{\langle K \rangle}k_{j}^{\alpha}})^{C}.
\end{eqnarray}
Assume a random variable $Y$ that represents the number of neighbor nodes infected by node $a$. Then the average value of $Y$ is:
\begin{eqnarray}
\mathrm{E}(Y) \nonumber  	
&=&\sum_{i=1}^{\langle K \rangle}\mathrm{E}(X_i) \\ 
&=& \langle K \rangle-\sum_{i=1}^{\langle K \rangle}(1-\frac{k_{i}^{\alpha}}{\sum_{j=1}^{\langle K \rangle}k_{j}^{\alpha}})^{C}.
\end{eqnarray}
Where the sum is over all the $\langle K \rangle$ neighbor nodes of  node $a$. However, in the epidemic spreading process the neighbor nodes of node $a$ may not be only susceptible nodes. To effevtively estimate the number of neighbor nodes that node $a$ infects, we need to know the number of susceptible nodes among all the neighbor nodes of node $a$. To estimate the total number of new infected nodes at time $t$, we count the  ratio $\mu$ of susceptible nodes among neighbor nodes of infected nodes in the network,  which is as follows: 
\begin{eqnarray}
\mu \nonumber  	
&=& \frac{\sum_{j=1}^{I(t)}n_j}{\sum_{j=1}^{I(t)}k_j} \\
&\approx& \frac{\langle n \rangle}{\langle K \rangle}.
\end{eqnarray}
Where $n_j$ is the number of susceptible nodes among the neighbor nodes of an infected node $j$. $\langle n \rangle$ represents the average  number of susceptible nodes among all the neighbors of infected nodes at time $t$. 
According to Eq. 6 and Eq. 7,  the total new infected nodes at time $t$ is:
\begin{eqnarray}
I_{new}(t) \nonumber  	
&\approx& \mathrm{E}(Y)\ast \mu \ast I(t) \\
&\approx&( \langle K \rangle-\sum_{i=1}^{\langle K \rangle}(1-\frac{k_{i}^{\alpha}}{\sum_{j=1}^{\langle K \rangle}k_{j}^{\alpha}})^{C}) \ast \frac{\langle n \rangle}{\langle K \rangle} \ast I(t).
\end{eqnarray}
Combining  Eq. 1 with Eq. 8, we get the dynamics equations of our model as follows:
\begin{equation}
\left \{
\begin{array}{rl}
\frac{\mathrm{d}I}{\mathrm{d}t} &=( \langle K \rangle-\sum_{i=1}^{\langle K \rangle}(1-\frac{k_{i}^{\alpha}}{\sum_{j=1}^{\langle K \rangle}k_{j}^{\alpha}})^{C})\ast \frac{\langle n \rangle}{\langle K \rangle }\ast I(t)-\lambda I,\\
\frac{\mathrm{d}S}{\mathrm{d}t} &=-( \langle K \rangle-\sum_{i=1}^{\langle K \rangle}(1-\frac{k_{i}^{\alpha}}{\sum_{j=1}^{\langle K \rangle}k_{j}^{\alpha}})^{C})\ast \frac{\langle n \rangle}{\langle K \rangle } \ast I(t),\\
\frac{\mathrm{d}R}{\mathrm{d}t} &= \lambda I.\\
   \end{array}
\right.
\end{equation}
\begin{figure}
 \centering
\includegraphics[width=5in,height=3in]{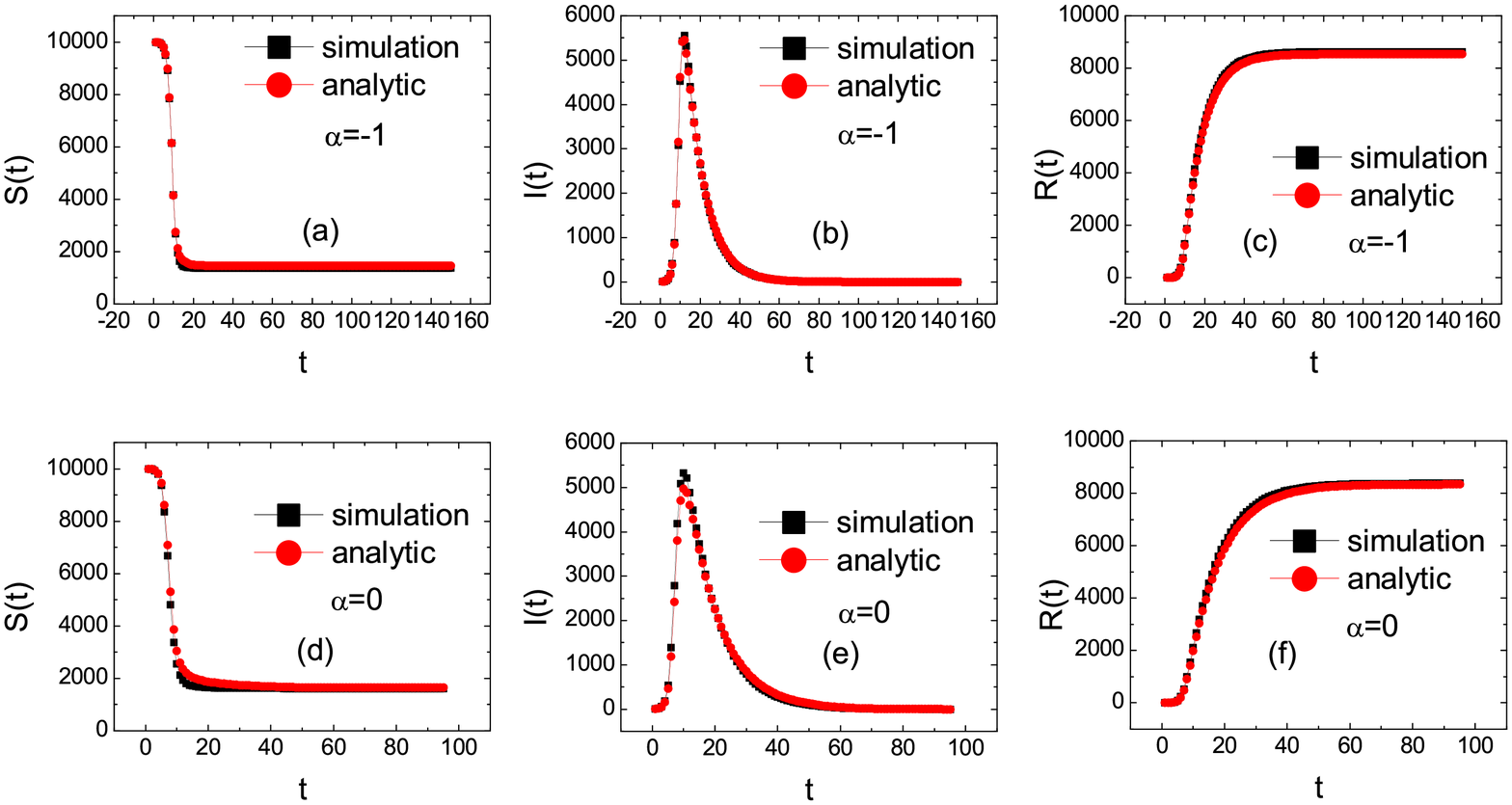}
\caption{$S(t)$, $I(t)$, and $R(t)$ vs. $ t$ for  $\alpha=0$ and $\alpha=-1$. A randomly selected node is set to be  infected initially. Delivery capacity of infected nodes is $C=5$.  $\lambda=0.1$. The network is generated by the static model\cite{Goh01}. The parameters are $N=10000$, $\langle k \rangle=5$, and $\gamma=2.5$.}
\end{figure}
We study the behaviours of  $S(t)$, $ I(t)$ and R(t) with increase of time $t$ on a large-scale scale-free network. In Fig. 2, $S(t)$ decreases abruptly, and then saturates with $t$.  $I(t)$  increases with $t$,  then decreases with $t$, and finally saturates. There is a peak of $I(t)$ that corresponds to the instantaneous maximum population of infected nodes. $R(t)$ increases abruptly, and then saturates with $ t$, which is opposite to $S(t)$. When $t$ is large enough, the epidemic spreading process stops, and $R(t)$ is  number of all the nodes that have ever been infected and removed finally from the disease. The trends of the curves for biased random walks of $\alpha=-1$ are similar with that of simple random walks of $\alpha=0$. 
 Also, the simulation results and the analytical results obtained from  Eq. 9 are consistent, as shown in Fig. 2.
\begin{figure}
 \centering
\includegraphics[width=5.2in,height=1.5in]{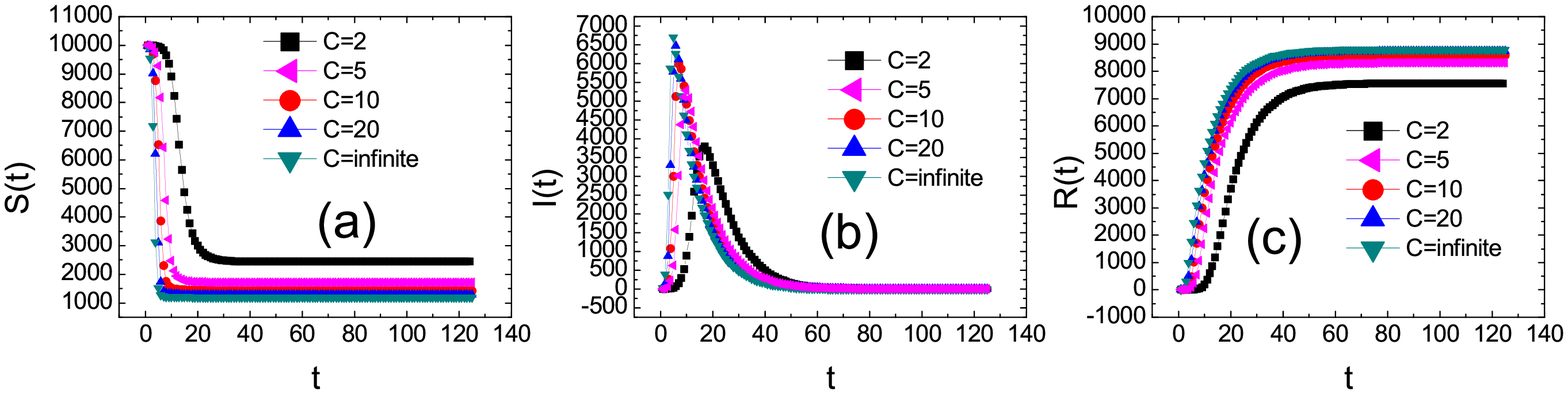}
\caption{$S(t)$, $I(t)$, and $R(t)$ vs. $ t$ for various delivery capacities C.   A randomly selected node is set to be  infected initially. $\alpha=0$.  $\lambda=0.1$. The network is generated by the static model. The parameters of the network are $N=10000$, $\langle k \rangle=5$, and $\gamma=2.5$.}
\end{figure}
\subsection{Factors of our model}
Delivery capacity $C$ of infected nodes is a critical factor in our epidemic spreading model. The larger the delivery capacity of an infected node, the more susceptible neighbor nodes an infected node will likely infects each time. When $C \to \infty$, $I_{new}(t) \to \langle n \rangle \ast I(t)$. Then Eq. 9 is reduced as follows:
\begin{equation}
\left \{
\begin{array}{rl}
\frac{\mathrm{d}I}{\mathrm{d}t} &=\langle n \rangle\ast I(t)-\lambda I,\\
\frac{\mathrm{d}S}{\mathrm{d}t} &=-\langle n \rangle\ast I(t),\\
\frac{\mathrm{d}R}{\mathrm{d}t} &= \lambda I.\\
\end{array}
\right.
\end{equation}
We show simulation results of $S(t)$, $I(t)$ and $R(t)$ for different value of $C$ in Fig.3. Clearly, the larger $C$, the faster  $S(t)$, $I(t)$ and $R(t)$ convergence,  and the epidemic spreads. 
The larger $C$, the larger population of nodes that has ever been infected which is inferred from S(t) and R(t) when $ t$ is large enough. The larger $C$, the larger instantaneous population of infected nodes   which is indicated from the peak of $I(t)$. 
\begin{figure}
 \centering
\includegraphics[width=5.2in,height=2.9in]{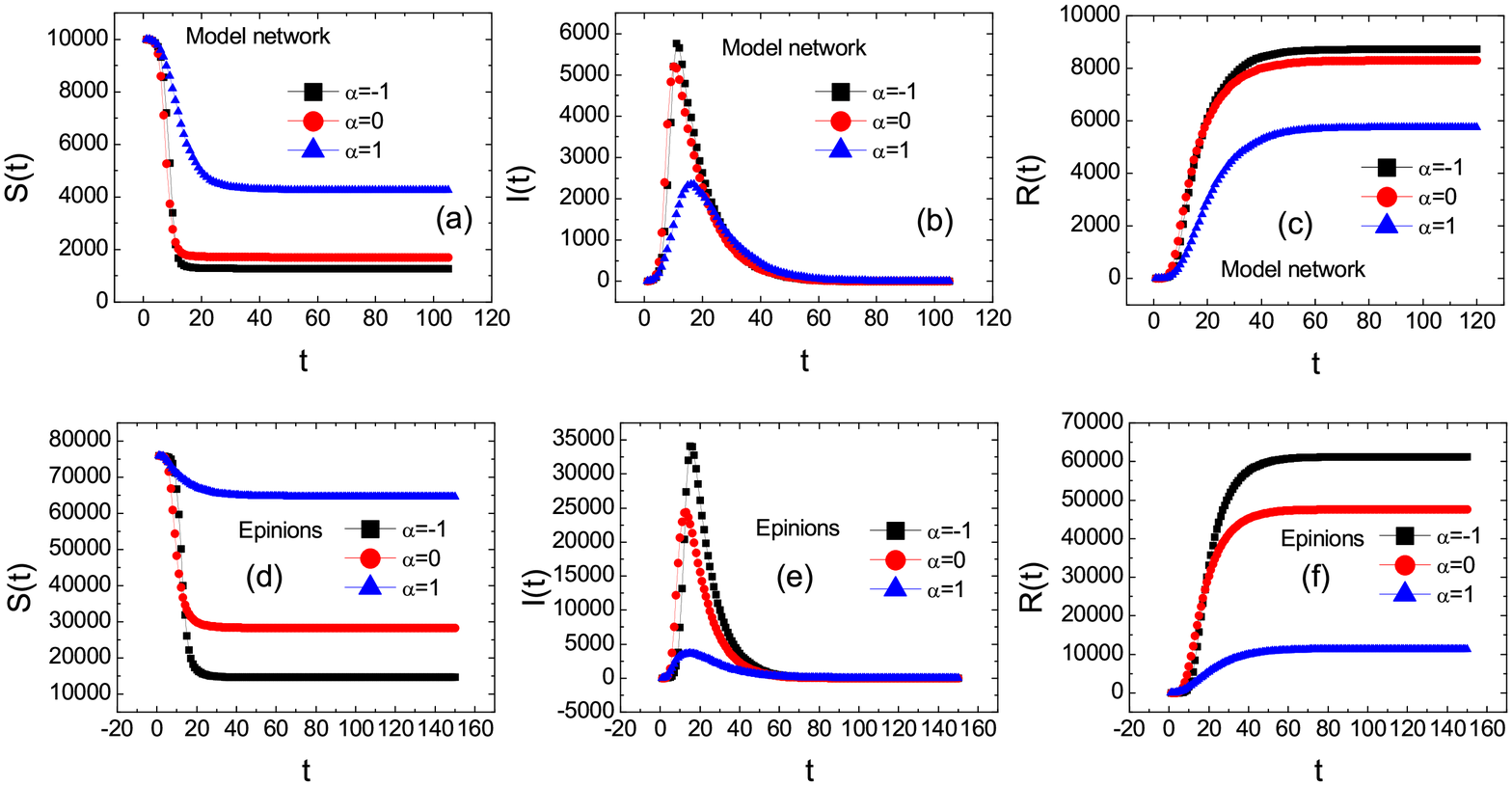}
\caption{$S(t)$, $I(t)$, and $R(t)$ vs. $ t$ for various $\alpha$.   A randomly selected node is set to be  infected initially. $C=5$.  $\lambda=0.1$. The network for (a), (b) and (c) is generated by the static model. The parameters of the network are $N=10000$, $\langle k \rangle=5$, and $\gamma=2.5$. The Epinions network for (d), (e) and (f) has  75, 879 nodes and 508, 960 edges.}
\end{figure}
The parameter $\alpha$ is another key factor in our epidemic model which determines the probability that the neighbors of an infected node get infected when $C$ is a limited constant. In Fig. 4, we see that $\alpha$ is correlated with the instantaneous maximum number of infected nodes $I_{peak}$, and the range of the spread that is reflected by the ultimate number of removed nodes $R_{end}$. 
 $\alpha=-1$ corresponds to a larger  $I_{peak}$ and a larger $R_{end}$  than $\alpha=0$ and $\alpha =1$. This indicates that random walks biased on small-degree nodes favors the epidemic spreading which is hold for the model network and the Epinions network, as shown in Fig.4. Then we investigate the optimal parameters $\alpha_{opt}$ that lead to the maximum  $I_{peak}$ and the maximum $R_{end}$  on the model network and the Epinions network. In Fig.5, we see $I_{peak}$ and $R_{end}$ as a function of $\alpha$. Clearly, $I_{peak}$ and $R_{end}$ increase with $\alpha$ first, then decrease with $\alpha$ respectively. There are $\alpha_{opt}$ that correspond to maximum $I_{peak}$ and maximum $R_{end}$ respectively. We  present the results of maximum $I_{peak}$, maximum $R_{end}$, and the corresponding $\alpha_{opt}$ for some real-world networks, as shown in Table 1.
\begin{table}[htbp]
\caption{\label{tab1}maximum $I_{peak}$ and maximum $R_{end}$ with corresponding optimal parameters $\alpha$ for real-world networks. $C=5$.  $\lambda=0.1$.}
\centering
\begin{tabular}{|c|c|c|c|c|c|c|}
\hline
NAME&NODES&EDGES&$I_{peak}$&$\alpha_{opt}$&$R_{end}$&$\alpha_{opt}$\\
\hline
Oregon-1&10790&22469&2903.33&-0.4&6010.37&-0.8\\
Gnutella &62586&147892&37833.43&-0.8&58155.71&-1.4\\
Epinions&75879&508837&35274.13&-0.8&61560.15&-1.2\\
Wiki-Vote&7115&103689&4294.22&-1&6623.23&-1.4\\
Yeast&2361&7182&1314.21&-0.8&2214.63&-1.2\\
email-Enron&36692&183831&12789.5&-0.8&24254.25&-1.4\\
Facebook&4039&88234&2105.11&-0.8&3920.21&-0.4\\
Geom&7343&11898&1778.5&-0.4&3202.5&-1\\
Political blogs&1222 &19021&813.26&-1.2&1192.14&-2.2\\
Power grid&4941&6594&1278.36&0.4&4304.27&-0.2\\
\hline
\end{tabular}
\end{table}

\begin{figure}
 \centering
\includegraphics[width=5in,height=2.1in]{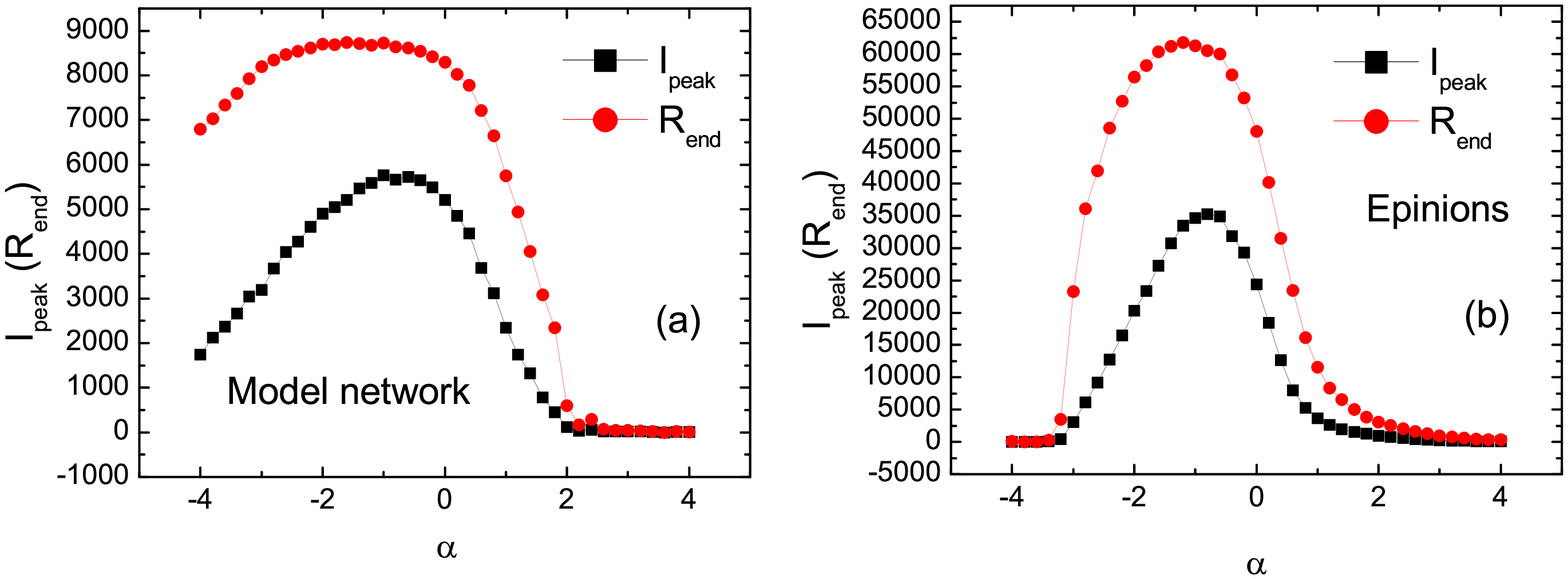}
\caption{ $I_{peak}$ and $R_{end}$ vs. $\alpha$.  A randomly selected node is set to be  infected initially.  $C=5$.  $\lambda=0.1$. The networks are the same as in Fig. 4. The results are the average of 100 independent runs. }
\end{figure}
\section{Impacts of networks structures on our model}
We investigate the influence of topological properties of complex networks including average node degree and degree distribution, on the behaviors of our epidemic spreading model. We focus on the spontaneous  number of infected nodes $I_{peak}$ and the final population of nodes that have ever been infected $R_{end}$, as well as the related optimal parameters $\alpha_{opt}$.
According Eq. 6, we have:
\begin{eqnarray}
\mathrm{E}(Y) \nonumber  	
&=& \langle K \rangle-\sum_{i=1}^{\langle K \rangle}(1-\frac{k_{i}^{\alpha}}{\sum_{j=1}^{\langle K \rangle}k_{j}^{\alpha}})^{C} \\ \nonumber
&=&  \langle K \rangle-\sum_{i=1}^{\langle K \rangle}(1- C\frac{k_{i}^{\alpha}}{\sum_{j=1}^{\langle K \rangle}k_{j}^{\alpha}}+\frac{C(C-1)}{2}(\frac{k_{i}^{\alpha}}{\sum_{j=1}^{\langle K \rangle}k_{j}^{\alpha}})^2-\cdots)\\
&=&  \sum_{i=1}^{\langle K \rangle}(C\frac{k_{i}^{\alpha}}{\sum_{j=1}^{\langle K \rangle}k_{j}^{\alpha}}-\frac{C(C-1)}{2}(\frac{k_{i}^{\alpha}}{\sum_{j=1}^{\langle K \rangle}k_{j}^{\alpha}})^2+\cdots).
\end{eqnarray}
When $\langle K \rangle \to \infty$, we get:
\begin{eqnarray}
\mathrm{E}(Y) \nonumber  	
&\approx& \sum_{i=1}^{\langle K \rangle}\frac{Ck_{i}^{\alpha}}{\sum_{j=1}^{\langle K \rangle}k_{j}^{\alpha}} \\
&\approx&  C
\end{eqnarray}
Eq. 12 means that generally the number of nodes that an infected nodes infects in one time step increases with average degree $\langle K \rangle$, and tends to $C$, which is further confirmed in Fig. 6. Also, for the whole epidemic spreading process,  the maximum $I_{peak}$ and the maximum $R_{end}$ increase substantially,  then saturate with $\langle K \rangle$ respectively, as shown in Fig. 7 and 8. 
Their corresponding optimal parameters $\alpha_{opt}$ are generally negative and decrease with $\langle K \rangle$.  This indicates that when the networks become dense, random walks should be more biased on small-degree nodes to make the epidemic spreading more intense and wide.
These results are consistent both for random networks (Fig. 7) and scale-free networks (Fig. 8). 
\begin{figure}
 \centering
\includegraphics[width=2in,height=2in]{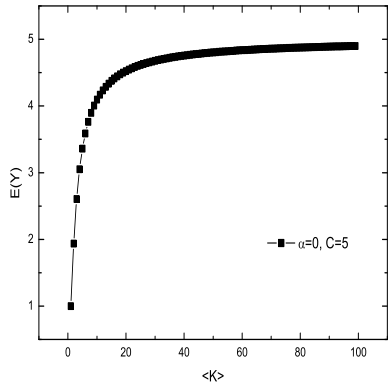}
\caption{ $\mathrm{E}(Y)$ vs. $\langle K \rangle$ obtained from Eq. 6.  }
\end{figure}
\begin{figure}
 \centering
\includegraphics[width=4in,height=2.7in]{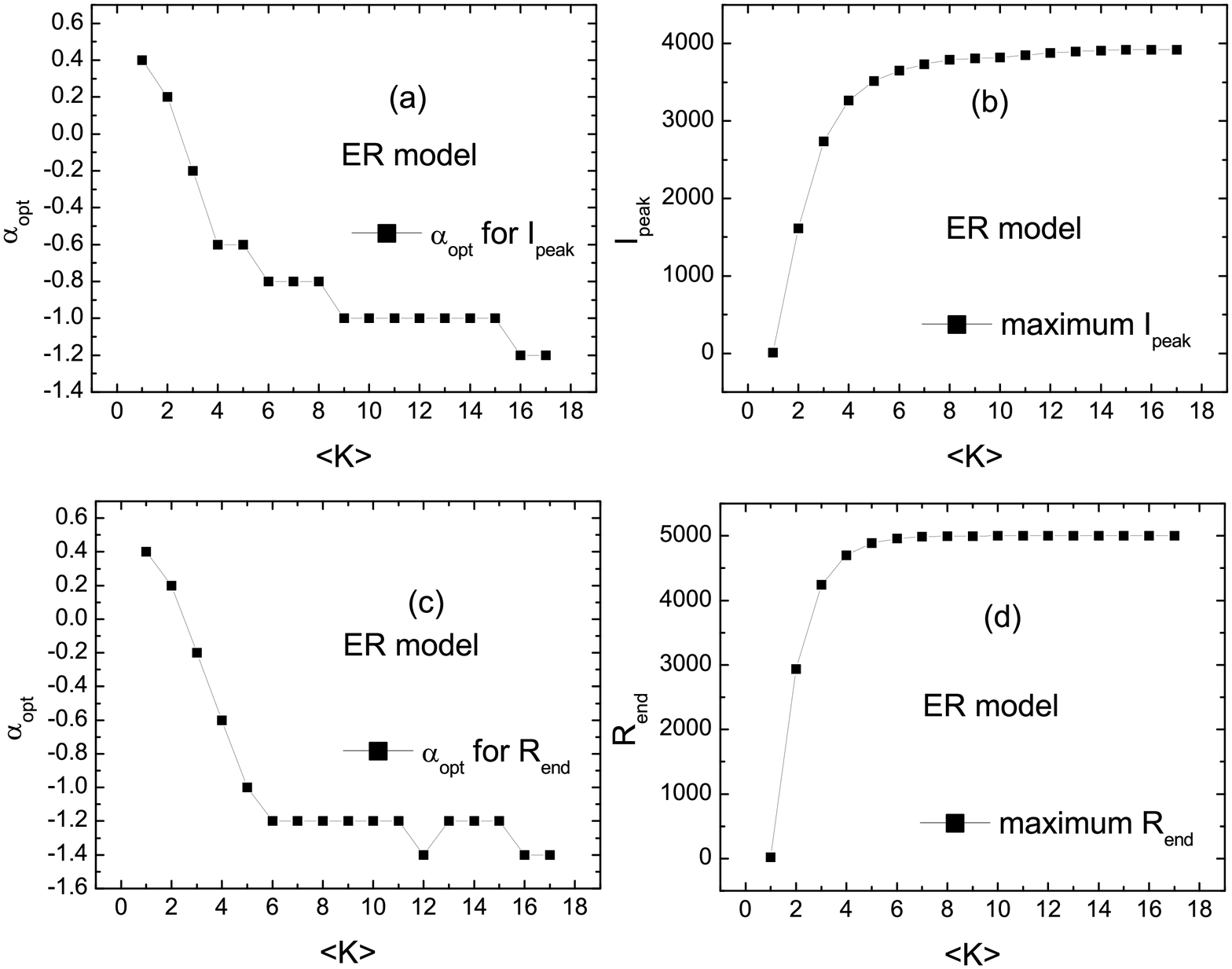}
\caption{ $\alpha_{opt}$ and the corresponding $I_{peak}$ and $R_{end}$   vs. $\langle K \rangle$ for random networks.  The networks are generated by the ER  (Erd\"{o}s-R\'{e}nyi) model\cite{Erdos}, and the  network size is $N=5000$. A randomly selected node is set to be  infected initially.  $C=5$.  $\lambda=0.1$. The results are the average of $ 10^4 $ independent runs. }
\end{figure}
\begin{figure}
 \centering
\includegraphics[width=4in,height=2.7in]{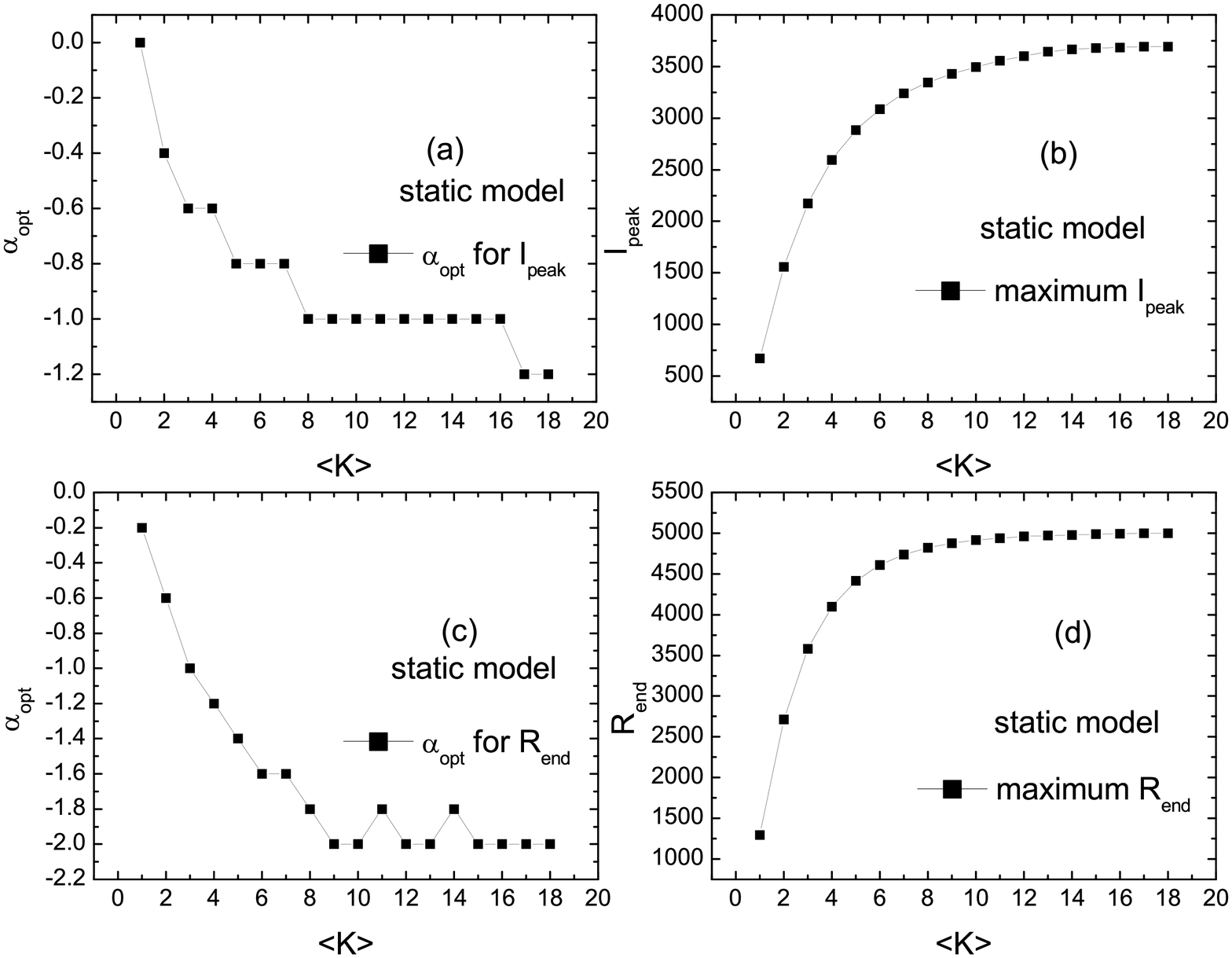}
\caption{ $\alpha_{opt}$ and the corresponding $I_{peak}$ and $R_{end}$   vs. $\langle K \rangle$ for scale-free networks.  The networks are generated by the static  model, the network size is $N=5000$, and the  power-law parameter is $\gamma=2.5$. A randomly selected node is set to be  infected initially.  $C=5$.  $\lambda=0.1$. The results are the average of $10^4$  independent runs. }
\end{figure}
We also investigate the impact of degree distribution on the epidemic spreading dynamics. In Fig. 9, the maximum $I_{peak}$ and the maximum $R_{end}$ increase abruptly,  then saturate with $\gamma$ respectively, and this means when the degree distribution becomes homogeneous, the spread of the epidemic becomes more fierce and wide in the network. The optimal parameters for $I_{peak}$ and $R_{end}$ increase with  $\gamma$.  This indicates when the network becomes homogeneous,  the extent that random walks are biased on small-degree nodes to get a wide and fierce epidemic spreading decreases.
 need less However, the fluctuations in the curves are clear.
\begin{figure}
 \centering
\includegraphics[width=4in,height=2.7in]{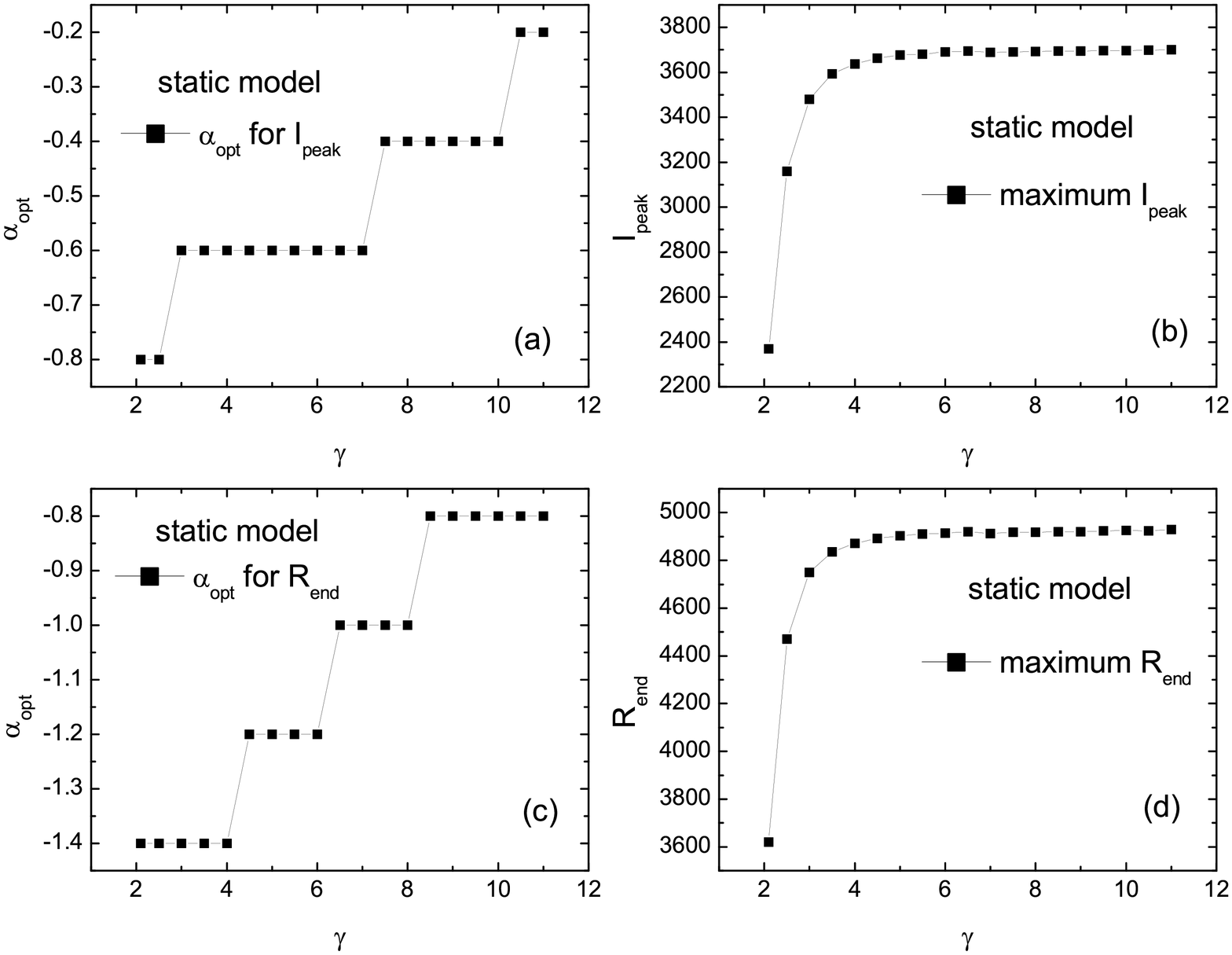}
\caption{  $\alpha_{opt}$ and the corresponding $I_{peak}$ and $R_{end}$   vs. power-law parameter $\gamma$.  The networks are generated by the static model, the network size is $N=5000$, and the average node degree is $\langle K \rangle=5$. A randomly selected node is set to be  infected initially.  $C=10$. $\lambda=0.1$.   The results are the average of  $10^4$  independent runs. }
\end{figure}
\section{Conclusions and discussions}
In summary, we investigate the epidemic spreading on complex networks including model networks and real-world networks. In our model, the epidemic spreading goes with packets transmission driven by biased random walks. Analytical and simulation results demonstrate that
Epidemic spreading becomes fierce and wide with increase of delivery capacity of infected nodes, average node degree, and the homogeneity of the network. The optimal parameters of the biased random walks in epidemic spreading are generally negative values. This means the random walks are biased on small-degree nodes to make an intense and wide spread of the epidemic. However, the biased random walks are based on only degrees of the nearest neighbor nodes in our model. The effects of biased random walks with more topological information on the epidemic spreading still need to be explored.
\section*{Acknowledgments}
This work was  supported by the Natural Science Foundation of China (Grant No. 61304154), the Specialized Research Fund for the Doctoral Program of Higher Education of China  (Grant No. 20133219120032), and the Postdoctoral Science Foundation of China (Grant No. 2013M541673).
%% The Appendices part is started with the command \appendix;
%% appendix sections are then done as normal sections
%% \appendix

%% \section{}
%% \label{}

%% If you have bibdatabase file and want bibtex to generate the
%% bibitems, please use
%%
%%  \bibliographystyle{elsarticle-num} 
%%  \bibliography{<your bibdatabase>}

%% else use the following coding to input the bibitems directly in the
%% TeX file.
%\cite{Jeong00}

\end{document}